\documentclass[[sigconf]{acmart}
\usepackage{amsmath}
\usepackage{algorithmic}
\usepackage{graphicx}
\usepackage{textcomp}
\usepackage{booktabs}
\usepackage{multirow}
\usepackage{subfigure}
\usepackage{xcolor}
\usepackage{enumitem}

\usepackage{fancyhdr}
\fancypagestyle{noheader}{
  \fancyhf{}% Clear header/footer
}

\usepackage{hyperref}
\usepackage{comment}

\AtBeginDocument{%
  \providecommand\BibTeX{{%
    \normalfont B\kern-0.5em{\scshape i\kern-0.25em b}\kern-0.8em\TeX}}}

\copyrightyear{2023} 
\acmYear{2023} 
\setcopyright{acmlicensed}
\acmConference[Accepted to ICIS'23]{}{December 10--13, 2023}{Hyderabad, India}
\newcommand{\bj}[1]{{\color{black}{#1}}}

\def\eg{\textit{e.g.}}
\def\ie{\textit{i.e.}}

\def\etc{\textit{etc.}}
\begin{document}

\sloppy
%%
%% The "title" command has an optional parameter,
%% allowing the author to define a "short title" to be used in page headers.
%\title{To Improve Is to Change: Mood Prediction Improves on Learning Emotion Change}
\title{A Virtual Reality Game to Improve Physical and Cognitive Acuity}
\renewcommand{\shorttitle}{VR Game for Physical and Cognitive Acuity}

%%
%% The "author" command and its associated commands are used to define
%% the authors and their affiliations.
%% Of note is the shared affiliation of the first two authors, and the
%% "authornote" and "authornotemark" commands
%% used to denote shared contribution to the research.
% \author{Paper ID: 5193}
\author{Blooma John}
\affiliation{
 \institution{Faculty of Science \& Technology, University of Canberra}
 \streetaddress{Faculty of Science \& Technology}
 \city{Canberra}
 \state{ACT}
 \country{Australia}}
 \email{blooma.john@canberra.edu.au}

\author{Ramanathan Subramanian}
\affiliation{%
  \institution{Faculty of Science \& Technology, University of Canberra}
 % \streetaddress{Faculty of Science \& Technology}
 \city{Canberra}
  \state{ACT}
  \country{Australia}}
  \email{ram.subramanian@canberra.edu.au}
  
\author{Jayan Chirayath Kurian}
\affiliation{
  \institution{School of Computer Science, University of Technology Sydney}
 \streetaddress{Faculty of Science \& Technology}
 \city{Canberra}
  \state{ACT}
  \country{Australia}}
  \email{ram.subramanian@canberra.edu.au}

% \author{Roland Goecke}
% \affiliation{%
%   \institution{University of Canberra}
%   %\streetaddress{8600 Datapoint Drive}
%   %\city{San Antonio}
%   \state{ACT}
%   \country{Australia}}
%   %\postcode{78229}
% \email{roland.goecke@ieee.org}
% \renewcommand{\shortauthors}{S. Narayana, R. Subramanian, I. Radwan, R. Goecke}

%%
%% By default, the full list of authors will be used in the page
%% headers. Often, this list is too long, and will overlap
%% other information printed in the page headers. This command allows
%% the author to define a more concise list
%% of authors' names for this purpose.
%\renewcommand{\shortauthors}{Narayana et al.}

%%
%% The abstract is a short summary of the work to be presented in the
%% article.
\begin{abstract}
We present the Virtual Human Benchmark (VHB) game to evaluate and improve physical and cognitive acuity. VHB simulates in 3D the BATAK lightboard game, which is designed to improve physical reaction and hand-eye coordination, on the \textit{Oculus Rift} and \textit{Quest} headsets. The game comprises the \textit{reaction}, \textit{accumulator} and \textit{sequence} modes; \bj{along} with the \textit{reaction} and \textit{accumulator} modes which mimic BATAK functionalities, the \textit{sequence} mode involves the user repeating a sequence of illuminated targets with increasing complexity to train visual memory and cognitive processing. A first version of the game (VHB v1) was evaluated against the real-world BATAK by 20 users, and their feedback was utilized to improve game design and obtain a second version (VHB v2). Another study to evaluate VHB v2 was conducted with 20 users, whose results confirmed that the deign improvements enhanced game usability and user experience in multiple respects. Also, logging and visualization of performance data such as \textit{reaction time}, \textit{speed between targets} and \textit{completed sequence patterns} provides useful data for coaches/therapists monitoring sports/rehabilitation regimens. 
\end{abstract}

%%
%% The code below is generated by the tool at http://dl.acm.org/ccs.cfm.
%% Please copy and paste the code instead of the example below.
%%

%%
%% Keywords. The author(s) should pick words that accurately describe
%% the work being presented. Separate the keywords with commas.

%% A "teaser" image appears between the author and affiliation
%% information and the body of the document, and typically spans the
%% page.
\begin{teaserfigure}
\centering
  \includegraphics[height=4cm, width=\textwidth]{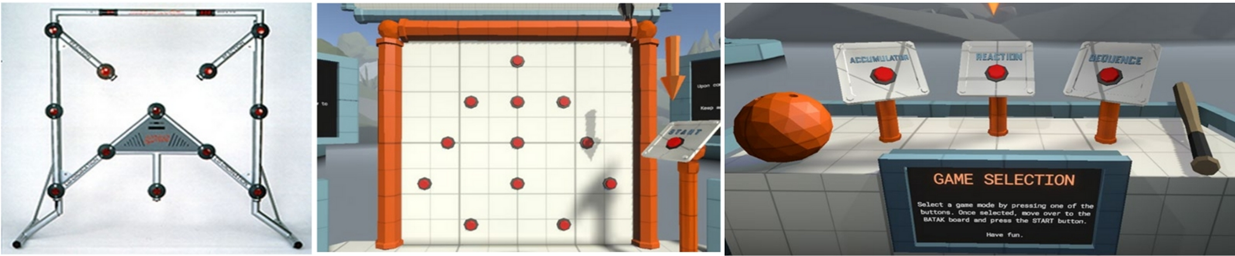}\vspace{-2mm}
  \caption{\textbf{Virtual Human Benchmark (VHB) Game:} We present a virtual simulation of the BATAK lightboard game (left) to benchmark and improve physical and cognitive performance. The VHB game (middle) is implemented on the \textit{Oculus Rift} and \textit{Quest} devices and offers users a 3D immersive experience. Users can activate either of the \textit{reaction}, \textit{accumulator} or \textit{sequence} game modes (right).}\vspace{-2mm}
\label{fig:VHB_overview}
\end{teaserfigure}

%%
%% This command processes the author and affiliation and title
%% information and builds the first part of the formatted document.
\maketitle

%
%-------------------------------------------
%

\section{Introduction}\label{sec:introduction}

Physical and mental health are both integral to healthy living and ageing, and a causal-cum-symbiotic relationship has been observed between the two~\cite{Ohrnberger16}. Motor skills are critical for performing everyday physical activities such as walking, running, \etc, for individuals and especially for athletes. Memory also represents a key cognitive skill that we employ on a daily basis for guiding our physical activities, and recalling and planning events. Brain regions responsible for motor functions and memory are highly vulnerable to age-related effects~\cite{SeidlerRachael2010}, and these skills can be severely impaired by disorders such as dementia, depression and stroke.  

Physical and cognitive activities such as exercise and board games~\cite{Altschul19} are known to promote healthy ageing and regularly feature in intervention activities at residential aged-care homes~\cite{Parekh18}. In this regard, the highly engaging BATAK lightboard game~\cite{batak10} known to improve hand-eye coordination, reflexes, motor skills, and stamina for sportspersons and individuals holds promise as a stimulating physical activity. Recently, video and virtual reality (VR) games~\cite{Cunha19,Cunha21} have also been found to be effective in engaging users, and improving dementia and stroke-related outcomes. Immersivity of VR games can transform mundane and repetitive exercise routines into competitive and stimulating experiences, and they can utilized by users to improve physical and cognitive performance from the comfort of their homes. 

\begin{figure}[!t]
\centering
\begin{minipage}{0.48\textwidth}
%\begin{figure}[t]
\centering
\includegraphics[width=\textwidth,height=4cm]{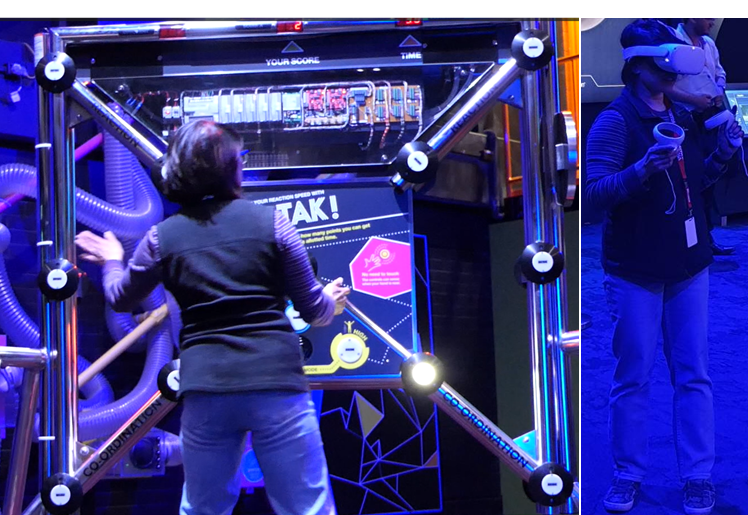}\vspace{-2mm}
\captionof{figure}{\textbf{Snapshots} of a user playing the BATAK game left), and the VHB game on the \textit{Oculus Quest} headset (right). Best-viewed under color and zoom.}\vspace{-3mm}
\label{fig:batak_vs_VHB}
\end{minipage} \hskip1em %
\begin{minipage}{0.48\textwidth}
\centering
\includegraphics[width=\textwidth,height=4cm]{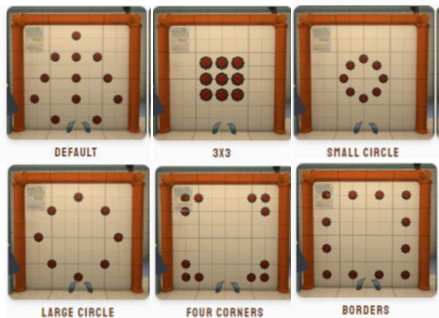}\vspace{-2mm}
\captionof{figure}{\textbf{Alternative lightboard configurations} to accommodate users with diverse requirements and abilities.}
\label{fig:alt_light_config}
\end{minipage}\vspace{-3mm}
\end{figure}

This work presents the Virtual Human Benchmark (VHB) VR game, which simulates the BATAK lightboard in a 3D immersive space. The VHB game works with the \textit{Oculus Rift} and \textit{Quest} headsets, and comprises the \textit{reaction}, \textit{accumulator} and \textit{sequence} game modes; while the \textit{reaction} and \textit{accumulator} modes behave similar to BATAK functionalities, the \textit{sequence} mode requires users to repeat a flashing light sequence thereby testing their visual memory. Overall, the VHB VR game can be utilized to train and improve one's physical and cognitive acuity. Fig.~\ref{fig:VHB_overview} presents snapshots of the BATAK and VHB games, while Fig.~\ref{fig:batak_vs_VHB} presents a user playing the real-word BATAK and VHB versions. 

Apart from game-play, the VHB gaming software also logs user performance data which can be visualized and analyzed via the \textit{VHB Insights} tool. Logged data include hand used for button press (left/right), time taken between button activations, hand distance from button over time, \etc\footnote{Exemplar game animations, user data logs and generated insights can be viewed at \url{https://virtual-human-benchmark.web.app/}}. Focus group interviews with health and sports experts revealed that the VHB game plus analytics can benefit sports training as well as physical and cognitive rehabilitation of people recovering from dementia and stroke. To evaluate VHB user experience, we performed a study where 20 users played both the BATAK and VHB games, and assessed them with respect to multiple attributes. VHB limitations inferred from the user study were addressed via improvements in game design, and a second user study confirmed that these improvements enhanced game usability and user experience. Overall, we make the following contributions:

\begin{itemize}[noitemsep,topsep=0pt]
\item[1.] We present the Virtual Human Benchmark, a virtual version of the BATAK lightboard game for training and improving physical and cognitive acuity. The game can be played from the comfort of the home using off-the-shelf \textit{Oculus Rift} and \textit{Quest} headsets. Also, the lightboard was made configurable to support users with varied requirements and abilities following recommendations from experts, who opined that the VHB game would benefit both the sports and health domains.
\item[2.] The gaming software also logs user performance data to generate analytics and visualizations conveying various measures. \bj{These measures} can provide actionable insights to sports coaches and \bj{rehabilitation }therapists. 
\item[3.] Two user studies were conducted to evaluate VHB usability and user experience. The first user study compared VHB against the real-world BATAK, and limitations with the VHB game were addressed via design improvements. The second study confirmed that design changes enhanced game usability and user experience.
\end{itemize}

\section{VHB Game Overview}
The objective of the Virtual Human Benchmark game is to create a set of challenges based on the BATAK lightboard game in the 3D VR space  to evaluate physical and cognitive acuity. The game modes are derived from the \textit{Human Benchmark} web application (\url{https://humanbenchmark.com/}) and the BATAK reaction test~\cite{batak10}. VHB is developed using the \textit{Unity} game engine, and other major software packages used include the \textit{Jetbrains Rider} integrated development environment, \textit{Oculus} software development kit (SDK), \textit{Universal Render Pipeline} for rendering shadows and textures, and the \textit{Windows Mixed Reality} SDK. The game has been successfully tested on the \textit{Oculus Rift}, \textit{Quest 2} and \textit{HP Reverb G2} VR headsets, and the minimum hardware specifications for an optimal development and testing experience, and game launching onto PC include: Nvidia GTX 970/AMD R9 290 GPU, Intel i5-4590 equivalent CPU, 8GB RAM, HDMI 1.3 video output, 3x USB 3.0 ports and 1x USB 2.0 port for the base station to track controller movements, and Windows 7 SP1 64-bit OS.

The VHB BATAK lightboard is shown in Fig.~\ref{fig:VHB_overview}(middle), while the menu for game mode selection via the \textit{Oculus} touch controllers is shown in Fig.~\ref{fig:VHB_overview}(right). The VHB game comprises the following challenge modes. 
\begin{itemize}[noitemsep,topsep=0pt]
\item \textbf{Reaction mode:} The VHB  lightboard will light all targets at random intervals, and the user reacts by pressing the controller button as fast as possible. This process repeats over a stipulated number of trials (typically 5 or 10), and the reaction time for each trial is recorded. This mode assesses the user’s ability to process visual information and physical reflexes.
\item \textbf{Accumulator mode:} Replicating the real-world BATAK functionality, the VHB lightboard will light up a single target button at random intervals, following which the user needs to strike the lit target quickly for another target to light up. This mode seeks to motivate the user to strike as many targets as possible during a set time limit of 30/60 seconds, and challenges the user’s motor agility, hand reach range and dexterity.
\item \textbf{Sequence mode:} While the reaction and accumulator modes primarily assess physical acuity, the sequence mode seeks to examine visual memory and cognitive processing. In this mode, the VHB lightboard will flash a sequence of targets one after another, which the user must repeat. The sequence pattern complexity will increase over trials until the user makes an error or the game completion following maximum number of trials.
\end{itemize}

To ensure that the VHB game seamlessly renders on multiple screens, a low-polygon design is adopted to design the scene model and assets (such as hands). To further enable the application to render at 75fps or higher frame rate so that the user feels immersed in a 3D environment, complementary colors like white, blue and orange are utilized for scene guidance. Blue color is used to denote non-interactive assets, while interactables such as the game buttons or mode selection menu are colored orange. An initial screen to inform and obtain user consent for data collection, is followed by the game selection menu. Upon selecting the desired game mode, the user will move towards the BATAK lightboard at the center-stage and press the `start' button to begin playing. A game manager module manages the game modes, controls how the BATAK lightboard buttons should behave in each mode, and how user data should be saved following play. A lightboard manager is also implemented to provide granular control relating to button activation, track activated buttons, and record whether the user presses the correct targets (for the accumulator and sequence modes). An walk-through of the three game modes is provided in the supplementary video.

\begin{figure*}[!t]
\centering
\includegraphics[height=4cm, width=\textwidth]{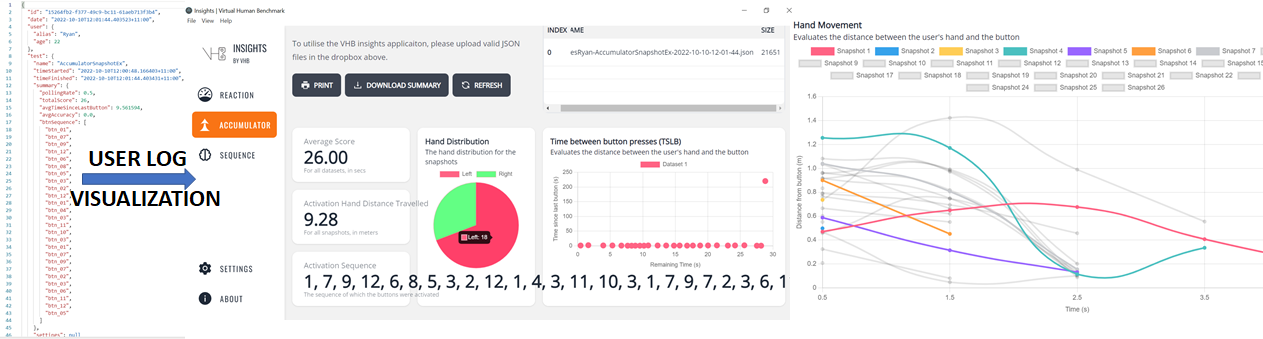}\vspace{-2mm} 
\caption{\textbf{User data analytics:} User performance logs in the form of \textit{achieved score}, \textit{buttons pressed} and their \textit{locations}, \textit{hand used} and \textit{distance covered} are stored in a json file (left), to generate visualizations via the insights tool (middle and right). These visualizations can provide actionable insights to allied stakeholders for charting training and rehabilitation regimens (best viewed under zoom).}\vspace{-4mm}
\label{fig:user_viz}
\end{figure*}

In terms of the data logged for each user, apart from summary data in the form of the score achieved and time taken, snapshots relating to each trial are also recorded (Fig.~\ref{fig:user_viz}). A \textit{reaction snapshot} includes reaction time and the time between successive lightboard flashes (random interval between 5--15s). The \textit{accumulator snapshot} records details such as the button position, time between two button presses, and remaining time, while key entries within a \textit{sequence snapshot} include the flashed sequence and sequence length, repeated sequence pattern, time taken to repeat pattern and time between target presses. The summary plus snapshot data per user are stored via \textit{json} files.

Figure~\ref{fig:user_viz} illustrates how the user performance data are used to generate visualizations. Feedback from health experts suggests that these analytics can provide actionable insights to allied stakeholders like sports coaches/therapists and clinicians utilizing VHB for training/rehabilitation purposes. The {json} \textit{accumulator} log (Fig.~\ref{fig:user_viz}(left)) is \bj{the input to} the \textit{VHB insights} tool to generate visualizations shown in Fig.~\ref{fig:user_viz}(middle) and Fig.~\ref{fig:user_viz}(right). Fig.~\ref{fig:user_viz}(middle) presents details like the user score, cumulative hand displacement (in meters), distribution of left/right hand usage as a pie-chart, button press sequence and a scatter plot showing remaining game time vs elapsed time between successive button presses (in seconds). We also utilize the hand-tracking capability of the \textit{Oculus} device to output the distance between the user's hand and target button over time Fig~\ref{fig:user_viz}(right). Hand tracking can be particularly useful for evaluating stroke rehabilitation outcomes, as stroke patients need repeated training to reach out for objects and grip them~\cite{Dean1997}. 

\subsection{Expert Feedback}
An initial version of the VHB game (VHB v1) was presented to a focus group of industry and academic experts for their evaluation and feedback. These included a VR specialist, a neuro-rehabilitation specialist, two sports health experts, a biomechanist and two researchers developing VR applications for health and well-being. Their comments are summarized as follows:
\begin{itemize}[noitemsep,topsep=0pt]
\item[1.] Experts conveyed positive impressions, and concurred that the VHB game could induce positive outcomes in the sports and health domains. Specifically, the VHB {reaction} and {accumulator} modes (and equivalently the BATAK lightboard game) can improve visual processing and visio-spatial skill which involves recognizing brightness/darkness, and is crucial for sports training and hand-eye coordination~\cite{Millard2021}. Additionally, the sequence mode induces cognitive stimulation, which is known to benefit patients with mild-to-moderate dementia~\cite{WoodsBob2012Csti}. 
\item[2.] A number of experts preferred to have a game version that supports seated users, and where the score was computed based on user ability and independent of time. One expert recommended a game mode where the user had to reach beyond 150\% arm's length. These recommendations were aimed at addressing requirements of specific demographics such as stroke patients, and training them to improve their sitting balance and object reaching abilities~\cite{Dean1997}. 
\item[3.] One expert also recommended the use of multiple lightboard layouts to support users with diverse requirements and abilities. All experts advocated the need for user experience surveys involving both gamers and non-gamers. 
\end{itemize}

In response to these recommendations, the game design was adapted to include multiple lightboard layouts as shown in Fig.~\ref{fig:alt_light_config}. The 3$\times$3 \textit{grid} and \textit{small circle} layouts were designed to support users with limited reach, while the \textit{large circle}, \textit{four corner} and \textit{border layouts} were intended to support users requiring training specific to upper limb strengthening and object reaching. We also proceeded to perform a user study to compare user experience with the real-world BATAK and the VHB counterpart as detailed below.

\section{User Study 1-- BATAK vs VHB v1 Comparison}
\begin{figure*}[t]
\centering
\includegraphics[height=3.6cm, width=\textwidth]{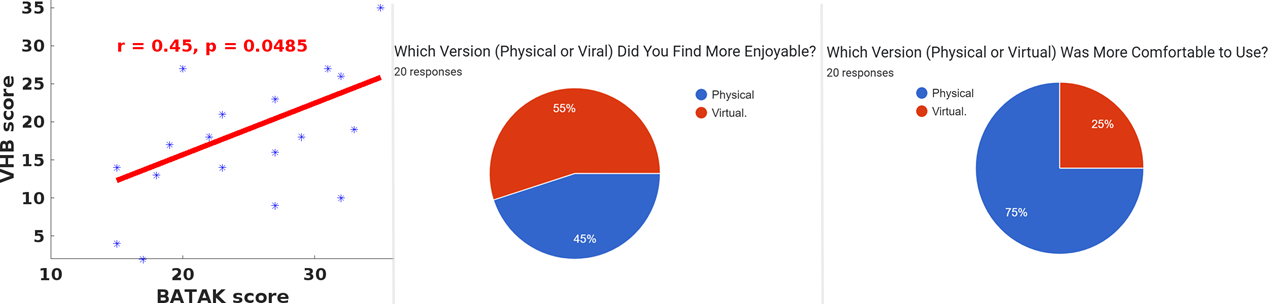}
\centering
\includegraphics[height=3.4cm, width=\textwidth]{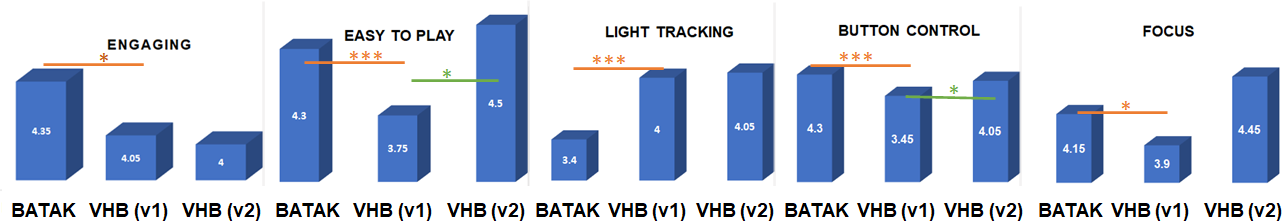}\vspace{-2mm}
\caption{\textbf{User study results summary:} (Top row, left-to-right) Scatter plot revealing a linear correlation between BATAK vs VHB user performance (Pearson $r=0.45$). Distribution of 20 user responses conveying which (physical/virtual) version was more \textit{enjoyable} and \textit{comfortable}. (Bottom row, left-to-right) Bar graphs depicting mean user scores for various game attributes. Orange lines and *'s denote significant BATAK vs VHB Version-1 score differences as given by a paired student $t$-test. Green lines and *'s denote significant VHB Version-1 vs Version-2 score differences conveyed by a two-sample $t$-test (* $\implies p<0.05, *** \implies p<0.001$).}\vspace{-6mm}
\label{fig:user_study}
\end{figure*}
\subsection{Objectives}\label{US1_obj}
To compare the VHB game experience against with the real-world BATAK lightboard, we performed a study with 20 healthy users at Questacon, Canberra. For a fair and exact comparison, users played the BATAK game and the VHB accumulator mode back-to-back, and filled a questionnaire soon after.  The specific aims of the study was to examine if (a) user performance was similar in the physical and virtual game versions (assuming similar levels of comfort with both), and (b) there were significant differences in user impressions with respect to game attributes such as \textit{engagement}, \textit{ease of play}, \textit{ease of light tracking}, \textit{simplicity of button press/control} (physically vs via the VR touch controller) and  \textit{level of focus/distraction} during game play in the physical vs virtual contexts (see Fig.~\ref{fig:user_study}(bottom row)). 

\subsection{Apparatus}\label{US1_app}
Eligible visitors, who were aged over 18 and had no physical disabilities, arriving at the BATAK board exhibit in Questacon were recruited for the study. Informed consent was obtained from these participants as per ethics approval, following which they were asked to additionally play the VHB accumulator game with the default lightboard layout (see Fig.~\ref{fig:VHB_overview}) using an \textit{Oculus Quest} headset. The BATAK and VHB lightboards identically comprised 12 targets. Figure~\ref{fig:batak_vs_VHB} presents an exemplar user playing the BATAK and VHB games within the same physical environment. The order of playing the virtual and real games was counterbalanced across participants. Upon completing both the games, users had to complete three online questionnaires; two concerning the physical and virtual game versions, and a third comparing the two. Users needed to provide both quantitative and qualitative responses in the game-specific questionnaires. Quantitative responses mainly involved specifying scores on a 1--5 Likert scale for the attributes specified in Section~\ref{US1_obj}, while qualitative responses denoted justifications for these scores. The comparison form involved selecting one of two alternate choices, plus explanations to support the selections. Including user preparation time, game play and questionnaire completion, each user took about 20 minutes to complete the study.

\subsection{Participants}
20 users (9 female, age range 31.2$\pm$ 13.7, 13 right-handed) took part in this study. 11 users reported that they frequently played video and VR games.% and one among them had experienced cybersickness during game play. 

\subsection{Results}
Fig.~\ref{fig:user_study} presents the user study results. We found a moderate and statistically significant Pearson correlation between the user BATAK and VHB scores (Fig.~\ref{fig:user_study} (top-left)), implying that the VHB game is ecologically valid, and agile users generally score well in both versions. Fig.~\ref{fig:user_study} (top-center, top-right) present a summary of the BATAK vs VHB comparisons, and convey that while users generally enjoyed playing both versions, the physical BATAK was more comfortable to play. Fig.~\ref{fig:user_study} (bottom row) summarizes mean user scores corresponding to multiple attributes of the BATAK and VHB (v1) games. The orange lines and *s denote significant score differences between the two, as conveyed by a paired $t$-test. The physical BATAK is perceived as being significantly better than VHB with respect to \textit{engagement}, \textit{ease of play}, \textit{button control} and \textit{extent of focus} (\ie, being less distracting), while the VHB version was seen to facilitate \textit{light tracking} much better. The pros and cons of the physical BATAK were further highlighted by qualitative comments, some of which are listed below:
\begin{itemize}[noitemsep,topsep=0pt]
\item \textit{It is easier than it looks, the field of view limits visibility of lit buttons in the corners of your vision.}
\item \textit{ I’m only 152cm tall and the lights at the extremities were harder to reach and spot via peripheral vision.}
\item \textit{The BATAK lightboard was quite good, however, I found it difficult to see the peripheral buttons when they lit up.}
\end{itemize}

The above remarks confirm that while the BATAK game play was facile, there were limitations with respect to its physical design and tracking of lights, especially those at the corners. Comments on the VHB game included:
\begin{itemize}[noitemsep,topsep=0pt]
\item \textit{Highly engaging, I found myself immersed to a point that it felt realistic and enjoyable.}
\item \textit{I found it more difficult to hit the buttons. The game wasn't as fluid as I thought it would be.}
\item \textit{Interesting, difficult at first mostly with pressing the virtual buttons.}
\item \textit{The experience was good. The application was well built, and easy to use, the contrast of a lit button was higher which made identification easier. The triangular pattern made for a more efficient eye scanning pattern as opposed to the physical BATAK test.}
\end{itemize}

These impressions convey that (a) the VHB game design was perceived to be immersive and realistic, and (b) the lightboard structure plus the illumination contrast made light tracking much easier. However, button activation via the touch controller was difficult, and considerably contributed to user frustration and distraction. These observations were further reinforced along with other issues such as weight of the VR headset in some comparative remarks listed below:
\begin{itemize}
[noitemsep,topsep=0pt]
\item \textit{I would rate the virtual version higher, as I found myself more immersed and had a higher quality experience.}
\item \textit{The physical board was more comfortable and easier to use.}
\item \textit{The virtual version was slightly less intuitive.}
\item \textit{Physical- No added weight on the head, the buttons were easier to press and more responsive.}
\end{itemize}

\subsection{Discussion of results and subsequent VHB design improvements}\label{US1_dis}
Overall, the user study revealed limitations of using VR devices and the VHB game design in particular. Discomfort with wearing the VR headset can be addressed via a more ergonomic VR device design and greater VR exposure. Usability issues with the VHB game were mainly noted with respect to executing physical movements such as button control; it needs to be noted here that while the VHB game can be simultaneously projected onto a monitor during game-play, the fact that the user study was conducted in an exhibit hall precluded this possibility. In the absence of any visual feedback, experimenters were only able to provide general guidance to users on executing game actions which was insufficient, and resulted in user distraction and frustration.

To better guide users to execute physical movements through game play, a second VHB game version (VHB v2) was developed with the following additional features:
\begin{itemize}[noitemsep,topsep=0pt]
\item \textbf{Tooltips:} Tooltips including a text panel succinctly instructing users on how to perform physical actions (\eg, activate buttons or walk through the scene) were added to provide visual feedback.
\item \textbf{Radial View Displays:} Adaptive animations that were rendered according to perspective (head pose) were created to demonstrate the game mode to users.
\item \textbf{3D Audio:} 3D spatial audio cues were incorporated to guide user attention to directions they are not facing. Examples include: helper arrows with a pulsating sound communicating to the user to select/start a game mode, audio feedback of button clicks and auditory notifications signifying game starting and completion.
\item \textbf{Haptic Feedback:} In addition to 3D audio, haptic feedback was implemented to signify the start and finish of a game mode, button activations, incorrect button presses and attempting to change game mode amidst play. The intensity and duration of feedback is variable depending on the performed action; \eg, incorrect actions trigger more pronounced feedback, whereas haptics accompanying valid button activations are light and short-lived.   
\end{itemize}

\section{User Study 2-- VHB v1 vs VHB v2 Comparison}
We then proceeded to evaluate user experience with the updated VHB version (VHB v2) via a second user study. 
\par{\textbf{Objective:}} The purpose of the study was to examine if user impressions and experience improved on incorporating visual, audio and haptic guidance features (Section~\ref{US1_dis}) in the VHB game design.

\par{\textbf{Apparatus:}} This study was conducted in a university lab, where users played the VHB (v2) accumulator game using an \textit{Oculus Quest} headset. They then completed the questionnaire specific to the virtual version in the first study. As in the first study, the  VR headset was not connected to a display monitor so that experimenters did not have any visual feedback to guide users; users were solely guided by visual, audio and haptic game features.   
\par{\textbf{Participants:}} 20 users (8 male, age range 33.8$\pm$9.1, 18 right-handed), not part of the first study, were recruited for the second upon providing informed consent. 
\par{\textbf{Results and Discussion:}} As users completed identical questionnaires in both user studies, we compared mean scores corresponding to VHB (v1) and VHB (v2) for the different game attributes, and the comparisons are depicted in Fig.~\ref{fig:user_study}(bottom row). Green lines and *s denote significant inter-version score differences, as conveyed by a two-sample $t$-test. While scores similar to VHB (v1) were observed for \textit{engagement} and \textit{ease of light tracking}, significantly higher scores for VHB (v2) were noted for the \textit{ease of play}, \textit{button control} and \textit{extent of focus} attributes. These results confirm that the incorporated features considerably enhanced game play, and consequently, user experience. Some qualitative user comments provided below are also reflective of these observations.

\begin{itemize}[noitemsep,topsep=0pt]
\item{\textit{I liked the simplicity of the game, it was very clear in what was required from the player and it was easy to play.}}
\item {\textit{Seemed well designed and immersive but generally feels more like a test than a game.}}
\item{\textit{It was clear and easy to manage, the recognition of hand movements was precise. The buttons were very low for me as 6ft, I had to bend down.}}
\item \textit{The overall experience has been good but would love to have more time in playing the game.}
\item \textit{Certainly immerses you and with future visual advancement (unreal engine), it will be life-like and will blur lines between real and virtual.}
\end{itemize}

Going forward, the VHB game design needs to be extended so that all three game modes involve multiple levels of difficulty, and can engage different demographics such as sportspersons, dementia and stroke patients to train tasks repetitively over long time-periods to achieve outcomes superior to traditional approaches. 

\section{Conclusion}
Expert opinions as well as qualitative and quantitative user opinions comparing the BATAK game with VHB (v1) revealed that the VHB game (a) can induce positive outcomes in the sports and health areas, (b) is engaging and (c) enables easier light tracking as compared to the physical BATAK. However, usability issues were noted, which hampered overall user experience. These issues were addressed via the implementation of visual, auditory and haptic user guidance features to obtain VHB (v2). A second study with an independent user group revealed that VHB (v2) usability and user experience were considerably superior to VHB (v1). Future work will involve tailoring the VHB game design to suit the requirements of specific demographics such as sportspersons, dementia and stroke patients, and validating with these communities if the VHB game achieves better user experience and physical/cognitive outcomes with respect to existing practices.

% \begin{acks}
% This research was supported partially by the Australian Government through the Australian Research Council's Discovery Projects funding scheme (project DP190101294).
% \end{acks}

%%
%% The next two lines define the bibliography style to be used, and
%% the bibliography file.

\bibliographystyle{ACM-Reference-Format}
\bibliography{references}
\thispagestyle{plain}
\end{document}